\documentclass[conference]{IEEEtran}
\usepackage[utf8]{inputenc}
\usepackage{array}
\usepackage{verbatim}
\usepackage{multirow}
\usepackage{amsmath}
\usepackage{amssymb}
\usepackage{graphicx}

\makeatletter

\providecommand{\tabularnewline}{\\}

\IEEEoverridecommandlockouts
\usepackage{cite}
\usepackage{amsfonts}\usepackage{algorithmic}
\usepackage{textcomp}
\usepackage{xcolor}

\DeclareMathOperator{\tr}{Tr}
\DeclareMathOperator{\detm}{det}
\newcommand{\herm}{^{{\dagger}}}

\graphicspath{ {./figures/} }
\def\BibTeX{{\rm B\kern-.05em{\sc i\kern-.025em b}\kern-.08em
    T\kern-.1667em\lower.7ex\hbox{E}\kern-.125emX}}

\makeatother

\begin{document}
\title{A closed-loop $2\times4$ downlink MIMO Framework for 5G New Radio
using OpenAirInterface}
\author{\IEEEauthorblockN{Duc Tung Bui and Le-Nam Tran} \IEEEauthorblockA{School of Electrical and Electronic Engineering, University College
Dublin, Ireland\\
 Email: duc.t.bui@ucdconnect.ie; nam.tran@ucd.ie}}
\maketitle
\begin{abstract}
We present the first-of-a-kind closed-loop $2\times4$ MIMO implementation
for the downlink of 5G Open RAN using OpenAirInterface (OAI), which
is capable of transmitting up to two transmission layers. Our implementation
is a fully functional 5G New Radio (5G NR) system, including the 5G
Core Network (5G CN), 5G Radio Access Network (5G RAN), as well as
5G NR User Equipment (UEs). This serves as a foundational framework
for further advancements in the context of emerging Open RAN (O-RAN)
development. A key feature of our implementation is the enhanced Channel
State Information (CSI) reporting procedure at the UE, which includes
Rank Indicator (RI), Precoding Matrix Indicator (PMI), and Channel
Quality Indicator (CQI). It is adjusted for the extended configuration
to maximize data rates. To demonstrate the performance of our implementation,
we measure the downlink data rates using \textit{iperf3} in two scenarios:
(i) fixed channels to assess two-layer data transmission and (ii)
\emph{Rice1} channels for general transmission analysis. The obtained
simulation results demonstrate that, compared to the existing $2\times2$
MIMO configuration in the OAI, our implementation improves the data
rates in almost all scenarios, especially at the high Signal-to-Noise-Ratios
(SNRs).
\end{abstract}


\section{Introduction}

Multiple-Input Multiple-Output (MIMO) technology serves as the cornerstone
of modern wireless communication systems, which allows for sending
and receiving multiple data signals simultaneously over the same radio
channel \cite{RN90}. In the context of fifth-generation mobile networks
(5G), the 3GPP has standardized closed-loop MIMO transmission schemes
that support up to 8 transmission layers \cite{RN171}, maximizing
the potential for spatial multiplexing and diversity gains. These
schemes rely on parameters related to channel state information (CSI)
reported by user equipment (UE), which enables the Base Station (gNB)
to optimize the transmission by determining the appropriate number
of transmission layers, selecting the proper modulation and coding
scheme (MCS), and choosing the most effective precoding matrix.

Historically, testing innovations and conducting practical experiments
in mobile cellular networks was largely exclusive to network operators
and vendors, due to the proprietary nature of the systems and strict
licensing restrictions. However, this landscape is evolving with the
advent of open source cellular stacks and affordable Software-Defined
Radio (SDR) devices. Several open source platforms now support 5G
Radio Access Network (RAN) simulation and deployment, including srsRAN
\cite{RN199,RN187,RN189,RN220}, OpenAirInterface5G (OAI 5G RAN) \cite{RN197,RN220,RN180},
and Open AI Cellular (OAIC) \cite{RN200,RN202}, the latter being
an expansion of srsRAN. After a thorough evaluation, we select the
OAI 5G RAN \cite{RN197} for our work due to its comprehensive feature
set and seamless integration with the OpenAirInterface 5G Core Network
(OAI 5G CN) \cite{RN201}.

OpenAirInterface (OAI) is an open source software project providing
a 3GPP-compliant implementation of key components for both 4G and
5G RANs, as well as the OAI 5G CN. The OAI 5G RAN and OAI 5G CN were
introduced and described in \cite{RN196}, which presented the preliminary
results available at the time of publication, along with a roadmap
for future development and an overview of OAI's roles in both research
and industrial contexts. Subsequent practical deployment scenarios
were explored in \cite{RN204,RN207,RN206}. The authors of \cite{RN195}
attempted to implement NR MIMO functions within OAI 5G RAN and verified
their implementation using the built-in rfsimulator. While they provided
detailed mathematical analyses, the downlink data rate was not presented.
In \cite{RN215}, a framework for Multi-User MIMO based on OAI was
demonstrated, along with detail analyses of the system's power and
energy consumption, as well as the provisional result of the uplink
data rate.

Currently, the OAI 5G RAN supports up to 4 transmitting (TX) antenna
at the gNB. However, its CSI reporting scheme is still limited to
the $2\times2$ configuration. This work develops a fully functional
5G NR system based on the OAI, and extends its CSI functionality.
Building upon the existing MIMO implementation, we present, a $2\times4$
closed-loop MIMO downlink communication scheme, in which the gNB is
equipped with up to 4 TX antennas and the UE with up to 2 receiving
(RX) antennas. To be specific, at the gNB, we propose and implement
a simple algorithm that enables the gNB to recognize the PMI for 1
and 2 transmission layers using 4 TX antennas. We also propose a new
Modulation and Coding Scheme (MCS) selection scheme based on the reported
CQI. On the UE side, we derive a new RI calculation scheme, as well
as a new PMI reporting procedure. Our implementation establishes a
framework for developing more advanced MIMO features within OAI, which
potentially leads to bigger impact on the evolution of the emerging
O-RAN.

The rest of this paper is organized as follows. Section. \ref{csirs_reporting}
provides an overview of the 5G NR downlink CSI reporting process and
outlines our contributions to enhancing MIMO communication capabilities
provided by OAI. Section \ref{subsec:Implementation} details our
system configuration and the evaluation setup. In Section \ref{subsec:Main-Results},
we present and discuss the results of the data rates achieved in various
scenarios. Finally Section. \ref{concl} concludes the paper.

\emph{Notation: }Bold face upper case letters are used for matrices.
The $\tr(\mathbf{X)}$, $\detm(\mathbf{X})$, and $\mathbf{X}\herm$
stand for the trace and determinant and Hermitian transpose of $\mathbf{X}$,
respectively.

\section{Proposed CSI Reporting Procedure for 5G NR Downlink $2\times4$ Closed-loop
MIMO \label{csirs_reporting}}

\begin{figure}[htbp]
\centerline{\includegraphics[width=0.95\columnwidth]{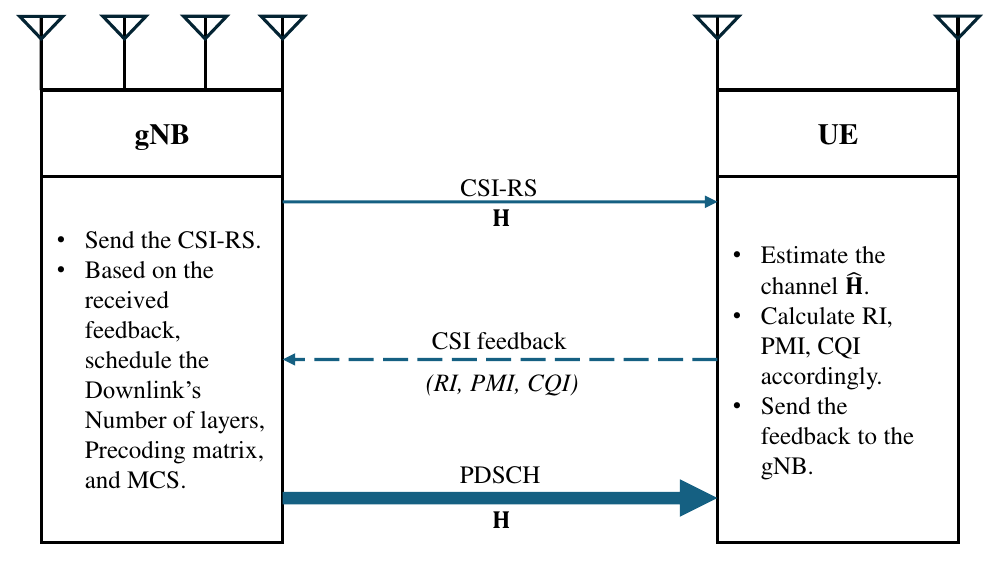}}
\caption{A visualization of the CSI reporting procedure}
\label{csirs}
\end{figure}

For closed-loop transmission, transmitters adapt their data transmission
according to the conditions of the wireless channel, which is termed
CSI in the context of 5G NR. In order to obtain the CSI, a gNB first
broadcasts a known reference signal, referred to as Channel State
Information Reference Signal (CSI-RS), to its UE. The UE uses the
received CSI-RS to estimate the channel and then reports the CSI parameters
to the gNB. For closed-loop MIMO transmission in 5G NR, the CSI feedback
generally includes three indicators: CQI, PMI, and RI. Based on the
CSI feedback, the gNB schedules downlink data transmissions via the
Physical Downlink Shared Channel (PDSCH), determining the number of
transmission layers, the precoding matrix, and the MCS. Fig. \ref{csirs}
illustrates the CSI reporting process. It is worth mentioning that
the CSI feedback only serves as a reference for the gNB, which is
not obligated to strictly follow it. However, in our implementation,
the gNB fully utilizes all three indicators - RI, PMI, and CQI - to
determine number of transmission layers, precoding matrix and MCS
accordingly.

\subsection{Rank Indicator Computation}

After estimating the wireless channel, the UE needs to compute and
report the RI to the gNB. The RI is proportional to the rank of the
wireless channel and reflects the potential for spatial multiplexing
gains. In other words, the RI provides information regarding the maximum
number of uncorrelated paths of the channel. This helps the gNB determine
the optimal number of transmission layers based on current channel
conditions. Other CSI indicators, such as PMI and CQI, are then computed
based on the RI. In our setup, we provide two possible configurations:
$2\times4$ MIMO and $1\times4$ MISO. Obviously, for the MISO case,
the RI is always 1. On the other hand, in the MIMO case, the RI can
be either 1 or 2. As mentioned previously, the current OAI's implementation
only supports calculating the RI for $2\times2$ channels. In this
work, we extend the RI calculation functionality so that the UE can
calculate the RI regardless the number of TX antennas at the gNB.
However, the number of the UE's RX antennas remains limited to 2,
as our focus is on the gNB.

The steps to calculate the RI are summarized as follows:
\begin{itemize}
\item For each subcarrier, based on the CSI-RS, the UE estimates the channel,
denoted by $\hat{\mathbf{H}}$. We adopt the channel estimation algorithm
provided by OAI.
\item Instead of directly computing the RI from $\hat{\mathbf{H}}$, the
UE defines the matrix $\mathbf{M}$ as:
\begin{equation}
\mathbf{M}=\hat{\mathbf{H}}\herm\hat{\mathbf{H}}
\end{equation}
 Note that the rank of $\mathbf{M}$ is also the rank of $\hat{\mathbf{H}}$.
It is known from MIMO theory, the capacity of the corresponding channel
depends of the eigenvalues of $\mathbf{M}$.
\item Since the UE has 2 antennas, $\mathbf{M}$ is always a $2\times2$
matrix, regardless of the number of antennas at the gNB. Next, the
UE calculates the following parameter: 
\begin{equation}
\gamma=\frac{\sum_{i,j}m_{ij}^{2}}{\detm\left(\mathbf{M}\right)}=\frac{m_{11}^{2}+m_{22}^{2}+m_{12}^{2}+m_{21}^{2}}{m_{11}m_{22}-m_{12}m_{21}}
\end{equation}
where $m_{ij}$ denotes the $(i,j)$-entry of $\mathbf{M}$. Let $\sigma_{1}$
and $\sigma_{2}$ denote the eigenvalues of $\mathbf{M}$. Then, $\gamma$
can be equivalently rewritten as
\begin{equation}
\gamma=\frac{\sum_{i,j}m_{ij}^{2}}{\detm\left(\mathbf{M}\right)}=\frac{\tr(\mathbf{M}\mathbf{M}\herm)}{\detm\left(\mathbf{M}\right)}=\frac{\sigma_{1}^{2}+\sigma_{2}^{2}}{\sigma_{1}\sigma_{2}}=\frac{\sigma_{1}}{\sigma_{2}}+\frac{\sigma_{2}}{\sigma_{1}}
\end{equation}
Note that $\gamma\geq2$.
\item It is easy to see that a large value of $\gamma$ implies a significant
difference between $\sigma_{1}$ and $\sigma_{2}$, suggesting that
$\hat{\mathbf{H}}$ is effectively rank-1 matrix. On the other hand,
if $\sigma_{1}$ and $\sigma_{2}$ are of similar values, making $\hat{\mathbf{H}}$
closer to a rank-2 matrix, $\gamma$ will approach $2$. Thus, we
define a threshold for $\gamma$, denoted by $\gamma_{th}$, to decide
if the involving channel is rank-1 or rank-2.
\item These steps are repeated for all subcarriers. If more subscarriers
exhibit rank-2 channels than rank-1, the RI is set to 2, and vice
versa.
\end{itemize}
After extensive testings, we find that $\gamma_{th}=2.5$ yields the
best performance in our considered scenarios.

\subsection{Precoding Matrix Indicator Selection}

While RI indicates the number of independent data streams the gNB
can schedule, the PMI guides the BS how to multiplex these streams
to optimize the rate. Specifically, PMI refers to an index that is
a suggestion for the gNB to select a proper Precoding Matrix from
a pre-determined codebook. Our implementation employs the Type I Single-Panel
Codebook, which is described in the Sect. 5.2.2.2.1 of \cite{RN171}.
In this configuration, the PMI includes $i_{2}$ and $i_{1}$, which
is further comprised of $i_{1,1}$, $i_{1,2}$, and $i_{1,3}$. The
parameter $i_{2}$ essentially indicates a coarse estimation of the
beam direction, while $i_{1}$ provides a finer estimate. In the current
OAI implementation, which supports CSI functionality for only 2 TX
antennas, there are four Precoding Matrices to select when $\textrm{RI}=1$,
and two when $\textrm{RI}=2$. Hence, $i_{1}$ is not necessary and
not reported.

The PMI computation becomes far more complex with 4 TX antennas at
the gNB. As mentioned earlier, RI can be either 1 or 2, and the PMI
depends on the value of RI. In both cases, there are 32 Precoding
Matrices for the gNB to select from the codebook.
\begin{itemize}
\item If $\textrm{RI}=1$, then $i_{2}$ and only $[i_{1,1}\;i_{1,2}]$
are reported. For 4 TX antennas, the possible values are $i_{1,1}=0,1,2,...,7$;
$i_{1,2}=0$; and $i_{2}=0,1,2,3$.
\item If $\textrm{RI}=2$, then $i_{2}$ and $[i_{1,1}\;i_{1,2}\;i_{1,3}]$
are reported. In this case, the values are $i_{1,1}=0,1,2,...,7$;
$i_{1,2}=0$; $i_{1,3}=0,1$; and $i_{2}=0,1$.
\end{itemize}
The process of determining the PMI (i.e. $i_{2}$ and $[i_{1,1}\;i_{1,2}\;i_{1,3}]$)
and the appropriate Precoding Matrix is as follows:
\begin{itemize}
\item After receiving the CSI-RS, the UE selects a proper codebook based
on the calculated RI.
\item The UE then computes the Signal-to-Interference-and-Noise-Ratio (SINR)
for each of the Precoding Matrices in the chosen codebook by summing
up the received signal power and the interference and noise power
across all the subcarriers. Note that, for the sake of optimizing
the speed and memory consumption of the software, the OAI quantifies
the calculated SINR directly into dB scale using a pre-determined
lookup table. As a result, the returned values of the SINR are always
integers.
\item Finally, the UE chooses the PMI corresponding to the Precoding matrix
that yields the highest SINR.
\end{itemize}

\subsection{Channel Quality Indicator Selection}

The final CSI parameter is the CQI, which is an integer between $0$
and $15$. It indicates channel quality and suggests the highest MCS
suitable for the downlink transmission to achieve the required block
error rate (BLER) under current channel conditions. The CQI is selected
based on the maximum SINR calculated in the PMI selection step, using
a lookup table that corresponds to a BLER of 0.1 or less. The lookup
table for the SINR--CQI mapping is shown in Table \ref{cqi_sinr},
which is constructed by from our extensive simulations. It is worth
mentioning that in our provided table, the CQI values start from 4
as our simulation results show that the transmission is still reliable
at this CQI value, even when the calculated SINR is low. Additionally,
when $\textrm{RI}=2$, the CQI is capped at 13, in order to guarantee
the reliability of the two layers the downlink transmission.
\begin{table}
\begin{centering}
\begin{tabular}{|c|c|c|}
\hline 
\multirow{2}{*}{\textbf{CQI}} & \multirow{2}{*}{\textbf{$\textrm{RI}=1$}} & \multirow{2}{*}{\textbf{$\textrm{RI}=2$}}\tabularnewline
 &  & \tabularnewline
\hline 
4 & $\textrm{SINR}\leq2$ & $\textrm{SINR}\leq2$\tabularnewline
\hline 
5 & $\textrm{SINR}=3$ & $\textrm{SINR}=3$\tabularnewline
\hline 
6 & $\textrm{SINR}\in\{4,5\}$ & $\textrm{SINR}\in\{4,5\}$\tabularnewline
\hline 
7 & $\textrm{SINR}\in\{6,7\}$ & $\textrm{SINR}\in\{6,7\}$\tabularnewline
\hline 
8 & $\textrm{SINR}\in\{8,9\}$ & $\textrm{SINR}\in\{8,9\}$\tabularnewline
\hline 
9 & $\textrm{SINR}=10$ & $\textrm{SINR}\in\{10,11\}$\tabularnewline
\hline 
10 & $\textrm{SINR}\in\{11,12\}$ & $\textrm{SINR}\in\{12,13\}$\tabularnewline
\hline 
11 & $\textrm{SINR}\in\{13,14,15\}$ & $\textrm{SINR}\in\{14,15\}$\tabularnewline
\hline 
12 & $\textrm{SINR}=16$ & $\textrm{SINR}\in\{16,\ldots,21\}$\tabularnewline
\hline 
13 & $\textrm{SINR}\in\{17,18\}$ & $\textrm{SINR}>21$\tabularnewline
\hline 
14 & $\textrm{SINR}=19$ & \multicolumn{1}{c}{}\tabularnewline
\cline{1-2}
15 & $\textrm{SINR}>19$ & \multicolumn{1}{c}{}\tabularnewline
\cline{1-2}
\multicolumn{1}{c}{} & \multicolumn{1}{c}{} & \multicolumn{1}{c}{}\tabularnewline
\end{tabular}
\par\end{centering}
\caption{Proposed lookup table to obtain CQI from SINR}
\label{cqi_sinr}
\end{table}

\section{Implementation and Simulation Results\label{sys_conf_title}}

\subsection{Implementation\label{subsec:Implementation}}

A fully functional OAI 5G NR system includes the 5G CN, 5G RAN, and
5G NR UEs. The 5G CN and 5G RAN are typically run on the same host
PC, however they can be configured to operate on different hosts as
disaggregating is a key feature of the O-RAN architecture, to which
the OAI is mainly dedicated. Using OAI's rfsimulator, UEs can be hosted
either on the same or on different hosts from the 5G CN and 5G RAN.
In this work we consider a single-user MIMO system but note that multiple
NR UEs can be served by a single gNB.

In order to minimize transmission line delays, we select an all-in-one
configuration for this work. The implemented system includes a gNB
which consists of the OAI 5G CN, the OAI 5G RAN, and a UE. The OAI
rfsimulator module supports various channel models. Again, to minimize
the delay due to channel modeling, which could result in significantly
lower calculated throughput compared to the true value, we adopted
the \textit{Rice1} model, which has just one tap. Additional taps
in the channel model increase computational complexity, potentially
extending the total transmission time and reducing the measured data
rate.

We configure the gNB and the UE to operate on the band \textit{n78},
which belongs to Frequency Range \textit{FR1} and uses the Time Division
duplex mode (TDD) \cite{RN203}. In term of bandwidth, the system
is set to utilize 106 resource blocks (RBs), with subcarrier spacing
of 30 kHz, corresponding to a bandwidth of $40$ MHz, including the
guardband. The 5G NR frame structure is described in detail in \cite{RN208}.

Once the connection between the UE and the gNB is established, one
can perform \textit{iperf3} to measure the data rate between the two
ends. In this work, we limited our consideration to the downlink performance
on the PDSCH using the TCP protocol. The data rate of the implemented
$4\times2$ implementation will be compared to the rate of the $2\times2$
configuration available within the OAI repository.

\subsection{Results and Discussions\label{subsec:Main-Results}}

\subsubsection{Fixed Channel}

To investigate the effect of transmitting data via 2 layers, we first
consider a fixed rank-2 wireless channel. The source code of the OAI
rfsimulator was adjusted to fix the values of the wireless channel
to a rank 2 matrix. In this way, the UE would always report an RI
of 2, which makes the gNB always transmit data using 2 layers. Specifically,
the $2\times4$ channel matrix is fixed to: 
\begin{equation}
\mathbf{H}_{2\times4}=\left[\begin{array}{cccc}
1 & 0.5 & 0.25 & 0.125\\
0.125 & 0.25 & 0.5 & 1
\end{array}\right]\label{chan_2x4}
\end{equation}
For this fixed channel, we consider two scenarios. In the first scenario,
we investigate the data rate using different CQI values when the transmission
is noise free. The source code of the UE is modified to report all
possible CQI values. In the second one, we examine the performance
of our system when additive white Gaussian noise (AWGN) is added to
the received signal at the UE. The noise power is normalized relatively
to the received signal power.

\begin{figure}
\includegraphics[scale=0.8]{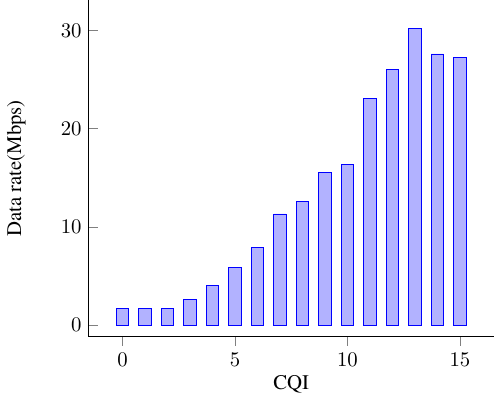}

\caption{Data rate versus the CQI value for noise-free transmission for the
fixed channel in \eqref{chan_2x4}.}

\label{data_cqi_2}
\end{figure}

Fig. \ref{data_cqi_2} plots the data rates for different CQI values
where the transmission is noise free using the MIMO channel in \ref{chan_2x4}.
The CQI and MCS tables utilized are 5.2.2.1-2 and 5.1.3.1-1 from \cite{RN171},
respectively. According to these tables, the CQI values from 0 to
2 all result in MCS 0, resulting in the same data rate for these CQI
values, as shown in Fig. \ref{data_cqi_2}. We can also observe that
the data rate peaks at CQI 13 then declines afterward. The reason
is that when the CQI increases, the coding rate and the order modulation
are increased accordingly, while inter-layer interference is relatively
fixed. After some point, the coding rate becomes unable to recover
data reliably, leading to packet retransmissions and, consequently,
a reduction in data rates. We note that this observation also holds
for other $2\times4$ channels we experimented, which are omitted
for the sake of brevity. Based on this observation, we set the maximum
CQI value to 13 in Table \ref{cqi_sinr} when $\textrm{RI}=2$.

\begin{figure}
\includegraphics[scale=0.8]{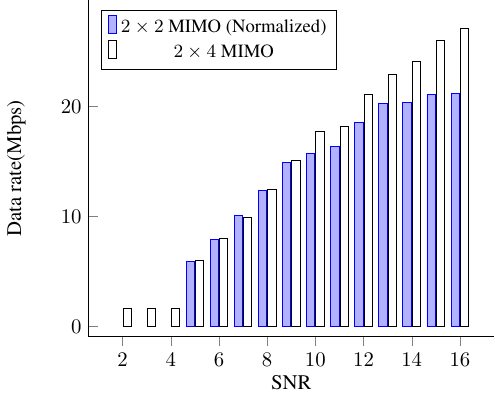}

\caption{Data rate versus SNR (i.e., normalized noise power) for the fixed
channel in \eqref{chan_2x4}.}
\label{data_noise_2}
\end{figure}

Next, Fig. \ref{data_noise_2} demonstrates the data rate versus SNR
(i.e. for noisy channels). It also includes the data rates for the
fixed $2\times2$ MIMO channel given by: 
\begin{equation}
\mathbf{H}_{2\times2}=\left[\begin{array}{cc}
1 & 0.5\\
0.5 & 1
\end{array}\right]\label{chan_2x2}
\end{equation}
We compare our $2\times4$ implementation to the $2\times2$ implementation
from the original OAI 5G RAN source code \cite{RN197}, but the rfsimulator's
source code was customized to always fix the simulated channel to
\ref{chan_2x2}. Comparing two systems with different numbers of antennas,
it is important to remark that the OAI rfsimulator takes twice more
time for the $2\times4$ MIMO case than the $2\times2$ MIMO case
since performing matrix multiplications on a $2\times4$ channel is
twice as complex as those for the $2\times2$ case. We also remark
that \textit{iperf3 }also includes the time spent by the rfsimulator
to calculating the achieved data rate. Thus, to obtain a fair comparison,
we halve the data rates for the $2\times2$ MIMO case. We can observe
from Fig. \ref{data_noise_2} that the data rate increases with SNR,
which is expected.The gains of the $2\times4$ MIMO systems over the
$2\times2$ counterpart are significant for high SNRs. Note also that
for the $2\times2$ MIMO case, the data rates for the first three
values associated with SNRs of 2, 3, and 4 are not shown. The reason
is that in the OAI's implementation, for these low SNRs, the UE reports
$\textrm{CQI}=0$ and the gNB automatically selects $\textrm{MCS}=9$,
which is very high for low SNRs. Indeed, in our experiment, we see
that this MCS results in a flood of re-transmissions and a high BLER.
As a result, the connection between the two ends is unreliable, and
the UE eventually stops working after some time.

\subsubsection{Random Channel}

\begin{figure}
\includegraphics[scale=0.8]{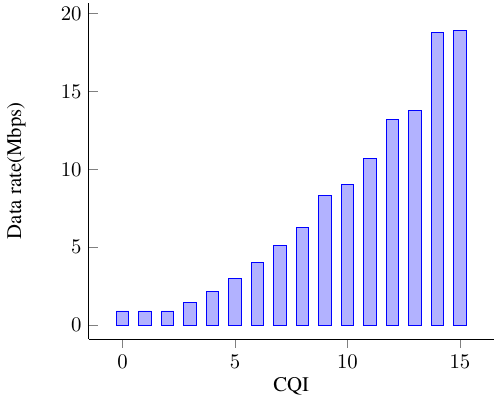}

\caption{Data rate versus the CQI value for noise-free transmission for the
fixed channel in for the \emph{Rice1} channel model.}
\label{data_cqi}
\end{figure}

In this subsection, we study the performance of the implemented $2\times4$
MIMO system using \emph{Rice1} channel model. We also consider two
scenarios, similar to the fixed channel cases described in the previous
subsection. Fig. \ref{data_cqi} plots the data rate for different
CQI values for the noise-free scenario. Again, the data rates for
CQI values of 0, 1, and 2 are identical since all correspond to the
same MCS. On the higher end, the data rates for CQI 14 and 15 are
nearly the same. This can be attributed to the fact that, at these
CQI values, the transmission is sensitive to the inter-layer interference.
We observe that for when $\textrm{CQI}=15$, re-transmissions occur
slightly more often than for $\textrm{CQI}=14$. This explains why
$\textrm{CQI}=15$ offers a minor improvement in data rate over $\textrm{CQI}=14$.
It would be reasonable to assume that if the CQI was allowed to increase,
the data rate would decrease for higher CQI values, similar to what
was observed in the previous section.

\begin{figure}
\includegraphics[scale=0.8]{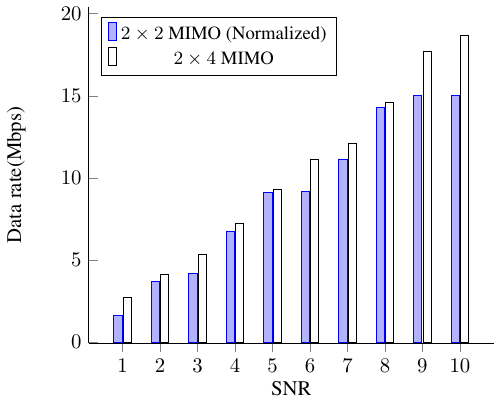}

\caption{Data rate versus SNR (i.e., normalized noise power) for the \emph{Rice1}
channel model.}
\label{data_noise}
\end{figure}

In the final numerical experiment, we study the performance of our
proposed system for noisy channels. Fig. \ref{data_noise} demonstrates
the data rate as a function of the SNR. Notably, when $\textrm{SNR}=10$
dB, (i.e. the normalized noise power is -10 dB), the data rate peaks
and equals the data rate of the noise-free channel where the highest
CQI is adopted (cf. Fig. \ref{data_cqi}). When compared to the $2\times2$
MIMO case, the $2\times4$ configuration consistently offers an improvement
in data rates, in particular at the high values of SNR.

\section{Conclusion and Future Work\label{concl}}

In this paper, we have implemented a closed-loop $2\times4$ MIMO
system based on the 5G RAN OAI, supporting up to 2 transmission layers
using Type I codebook. Specifically, a fully functioning 5G NR system,
which includes 5G CN, 5G RAN, and NR UEs was built for the first time,
which serves as a framework for further extensions or optimizations.
At the UE, we improved the RI calculation, extended the PMI reporting
procedure to support 4 TX antennas at the gNB,, and modified the CQI
reporting to maximize the data rates of 2-layer transmissions. The
simulation results have shown that the implemented $2\times4$ MIMO
system offers considerable improvements in terms of the data rates
in all the considered scenarios over the $2\times2$ MIMO counterpart
currently available on OAI, especially at the high SNRs.

For future work, our first priority is to replace the OAI rfsimulator
by real-world transmission using SDR devices at both the gNB and the
UE. We aim to report our results for real-world transmission in the
very near future. Additionally, we plan to extend the current single-user
$2\times4$ MIMO setup to support multi-user MIMO, which is the primary
motivation for this work.

\section*{Acknowledgment}

This publication has emanated from research supported by Taighde Éireann
- Research Ireland under Grant numbers 22/US/3847 and 13/RC/2077\_P2
at \emph{CONNECT: the Research Ireland Centre for Future Networks}.

\bibliographystyle{IEEEtran}
\bibliography{IEEEabrv,mybib}

\begin{thebibliography}{10}
\providecommand{\url}[1]{#1}
\csname url@samestyle\endcsname
\providecommand{\newblock}{\relax}
\providecommand{\bibinfo}[2]{#2}
\providecommand{\BIBentrySTDinterwordspacing}{\spaceskip=0pt\relax}
\providecommand{\BIBentryALTinterwordstretchfactor}{4}
\providecommand{\BIBentryALTinterwordspacing}{\spaceskip=\fontdimen2\font plus
\BIBentryALTinterwordstretchfactor\fontdimen3\font minus
  \fontdimen4\font\relax}
\providecommand{\BIBforeignlanguage}[2]{{%
\expandafter\ifx\csname l@#1\endcsname\relax
\typeout{** WARNING: IEEEtran.bst: No hyphenation pattern has been}%
\typeout{** loaded for the language `#1'. Using the pattern for}%
\typeout{** the default language instead.}%
\else
\language=\csname l@#1\endcsname
\fi
#2}}
\providecommand{\BIBdecl}{\relax}
\BIBdecl

\bibitem{RN90}
M.~K. Samimi, S.~Sun, and T.~S. Rappaport, ``{MIMO channel modeling and
  capacity analysis for 5G millimeter-wave wireless systems},'' in \emph{2016
  10th European Conference on Antennas and Propagation (EuCAP)}, Conference
  Proceedings, pp. 1--5.

\bibitem{RN171}
3GPP, ``{TS 38.214 - NR Physical layer procedures for data}.''

\bibitem{RN199}
\BIBentryALTinterwordspacing
``{srsRAN} {Project}.'' [Online]. Available: \url{https://www.srslte.com/5g}
\BIBentrySTDinterwordspacing

\bibitem{RN187}
L.~Mamushiane, A.~Lysko, H.~Kobo, and J.~Mwangama, ``{Deploying a Stable 5G SA
  Testbed Using srsRAN and Open5GS: UE Integration and Troubleshooting Towards
  Network Slicing},'' in \emph{2023 International Conference on Artificial
  Intelligence, Big Data, Computing and Data Communication Systems (icABCD)},
  Conference Proceedings, pp. 1--10.

\bibitem{RN189}
J.~E. Hakegard, H.~Lundkvist, A.~Rauniyar, and P.~Morris, ``{Performance
  Evaluation of an Open Source Implementation of a 5G Standalone Platform},''
  \emph{IEEE Access}, vol.~12, pp. 25\,809--25\,819, 2024.

\bibitem{RN220}
M.~Amini and C.~Rosenberg, ``{A Comparative Analysis of Open-Source Software in
  an E2E 5G Standalone Platform},'' in \emph{2024 IEEE Wireless Communications
  and Networking Conference (WCNC)}, Conference Proceedings, pp. 1--6.

\bibitem{RN197}
\BIBentryALTinterwordspacing
``{OpenAirInterface} {5G}.'' [Online]. Available:
  \url{https://gitlab.eurecom.fr/oai/openairinterface5g}
\BIBentrySTDinterwordspacing

\bibitem{RN180}
R.~Mundlamuri, S.~Badran, R.~Gangula, F.~Kaltenberger, J.~M. Jornet, and
  T.~Melodia, ``{5G Over Terahertz Using OpenAirInterface},'' in \emph{2024
  19th Wireless On-Demand Network Systems and Services Conference (WONS)},
  Conference Proceedings, pp. 29--32.

\bibitem{RN200}
\BIBentryALTinterwordspacing
``Open {AI} {Cellular}.'' [Online]. Available:
  \url{https://openaicellular.github.io/oaic/}
\BIBentrySTDinterwordspacing

\bibitem{RN202}
N.~H. Stephenson, A.~J. Chiejina, N.~B. Kabigting, and V.~K. Shah,
  ``{Demonstration of Closed Loop AI-Driven RAN Controllers Using O-RAN SDR
  Testbed},'' in \emph{MILCOM 2023 - 2023 IEEE Military Communications
  Conference (MILCOM)}, Conference Proceedings, pp. 241--242.

\bibitem{RN201}
\BIBentryALTinterwordspacing
``{OpenAirInterface} {CN}.'' [Online]. Available:
  \url{https://gitlab.eurecom.fr/oai/cn5g}
\BIBentrySTDinterwordspacing

\bibitem{RN196}
\BIBentryALTinterwordspacing
F.~Kaltenberger, A.~P. Silva, A.~Gosain, L.~Wang, and T.-T. Nguyen,
  ``{OpenAirInterface: Democratizing innovation in the 5G Era},''
  \emph{Computer Networks}, vol. 176, p. 107284, 2020. [Online]. Available:
  \url{https://www.sciencedirect.com/science/article/pii/S1389128619314410}
\BIBentrySTDinterwordspacing

\bibitem{RN204}
R.~M. Ursu, A.~Papa, and W.~Kellerer, ``{Experimental Evaluation of Downlink
  Scheduling Algorithms using OpenAirInterface},'' in \emph{2022 IEEE Wireless
  Communications and Networking Conference (WCNC)}, Conference Proceedings, pp.
  84--89.

\bibitem{RN207}
A.~Sahbafard, R.~Schmidt, F.~Kaltenberger, A.~Springer, and H.~P. Bernhard,
  ``{On the Performance of an Indoor Open-Source 5G Standalone Deployment},''
  in \emph{2023 IEEE Wireless Communications and Networking Conference (WCNC)},
  Conference Proceedings, pp. 1--6.

\bibitem{RN206}
T.~O. Atalay, A.~Famili, D.~Stojadinovic, and A.~Stavrou, ``{Demystifying 5G
  Traffic Patterns with an Indoor RAN Measurement Campaign},'' in \emph{IEEE
  GLOBECOM 2023}, Conference Proceedings, pp. 1185--1190.

\bibitem{RN195}
K.~A. Saaifan, T.~Schlichter, and T.~Heyn, ``{NR MIMO Feature Implementation
  into OpenAirInterface},'' in \emph{WSA 2021; 25th International ITG Workshop
  on Smart Antennas}, Conference Proceedings, pp. 1--6.

\bibitem{RN215}
G.~N. Katsaros, M.~Filo, R.~Tafazolli, and K.~Nikitopoulos, ``{MIMO-SoftiPHY: A
  Software-Based PHY Design and Implementation Framework for Highly-Efficient
  Open-RAN MIMO Radios},'' \emph{IEEE Transactions on Mobile Computing},
  vol.~23, no.~12, pp. 12\,491--12\,504, 2024.

\bibitem{RN203}
3GPP, ``{TS 38.101 - NR User Equipment (UE) radio transmission and reception;
  Part 1: Range 1 Standalone}.''

\bibitem{RN208}
------, ``{TS 38.211 - NR Physical channels and modulation}.''

\end{thebibliography}
 \vspace{12pt}

\end{document}